# Tungsten boride shields in a spherical tokamak

Colin G Windsor[1], Jack O Astbury[1], James Davidson[2], Charles J R McFadzean[2], J Guy Morgan[3], Christopher Wilson[1], Samuel A Humphry-Baker[2]

[1] Tokamak Energy Ltd, 173 Brook Drive, Milton Park, Oxon, OX14 4SD, UK

[2] Dept. of Materials, Imperial College, London SW7 2AZ, UK

[3] Culham Electromagnetics, D5, Culham Science Centre, Abingdon, OX14 3DB

**Abstract**

The favourable properties of tungsten borides for shielding the central High Temperature Superconductor (HTS) core of a spherical tokamak fusion power plant are modelled using the MCNP code. The objectives are to minimize the power deposition into the cooled HTS core, and to keep HTS radiation damage to acceptable levels by limiting the neutron and gamma fluxes. The shield materials compared are $W_2B$, $WB$, $W_2B_5$ and $WB_4$ along with a reactively sintered boride $B_{0.329}C_{0.074}Cr_{0.024}Fe_{0.274}W_{0.299}$, monolithic W and WC. Of all these $W_2B_5$ gave the most favourable results with a factor of ~10 or greater reduction in neutron flux and gamma energy deposition as compared to monolithic W. These results are compared with layered water-cooled shields, giving the result that the monolithic shields, with moderating boron, gave comparable neutron flux and power deposition, and (in the case of $W_2B_5$) even better performance. Good performance without water-coolant has advantages from a reactor safety perspective due to the risks associated with radio-activation of oxygen. $^{10}B$ isotope concentrations between 0 and 100% are considered for the boride shields. The naturally occurring 20% fraction gave much lower energy depositions than the 0% fraction, but the improvement largely saturated beyond 40%. Thermophysical properties of the candidate materials are discussed, in particular the thermal strain. To our knowledge, the performance of $W_2B_5$ is unrivalled by other monolithic shielding materials. This is partly as its trigonal crystal structure gives it higher atomic density compared with other borides. It is also suggested that its high performance depends on it having just high enough $^{10}B$ content to maintain a constant neutron energy spectrum across the shield.

Keywords: fusion, spherical tokamaks, tungsten borides, shielding, neutrons, gamma rays

## 1. Introduction

Spherical tokamaks present a unique opportunity to accelerate the delivery of safe, carbon-free, abundant, base-load fusion power [1]. They have the plasma closer to the current-carrying central column and so make more efficient use of the magnetic field, which decreases with distance from the column. Combining spherical tokamaks with the high field capabilities of High Temperature Superconductors (HTS) offer a potential route to a smaller power plant [2]. The low aspect ratio of the spherical tokamak presents a difficulty: for an HTS core of radius $R_{core}$ necessary to provide the magnetic field, the shield thickness available for a plasma of major radius $R_0$, aspect ratio $A$ and a vacuum gap $g$ between plasma and shield is $R_0(A-1)/A - R_{core} - g$. The lower the aspect ratio and major radius, the thinner the space available for a shield. Several studies of candidate shielding materials [3-8] have been made using the Monte Carlo modelling code MCNP [9]. It has been shown that the inclusion of boron within a tungsten-based shield is advantageous, with its high neutron absorption cross section at lower energies[4]. Various practical efforts to fabricate and test the properties of such materials have begun [10-13] but the search is on to find the optimal boron-containing material, and optimal parameters of a tokamak shield of specified lifetime.

    The precise constraints that any shield must satisfy are not yet known. The total power deposition determines the costs of the cryogenic plant needed to keep the HTS at operating temperatures of say 20 K. Calculations suggest that it would be a relatively low fraction of the cost of a 200 MW fusion power plant [3]. More difficult to calculate are the radiation damage constraints. Experimental measurements from HTS tapes irradiated in fission reactors [14] suggest a limit of order $2\times10^{18}$ cm$^{-2}$, but the energy spectrum of the flux from such reactors is different from that behind a tokamak shield, the temperatures are several hundred K higher, and the damage rates much faster. The present work focuses on the properties of tungsten borides as shield materials. Section 2 indicates why they have advantages. Section 3 details the MCNP methods used and the parameters calculated.

Section 4 compares several monolithic, homogeneous such materials. Section 5 presents results when water cooling is introduced. Section 6 examines the effect of varying the proportion of the favourable boron 10 isotope and finally Section 7 presents some thermophysical properties of the materials. Considerations of the activation of the materials, their decay after shutdown, loss of $^{10}$B isotope and radiation damage, all using the FISPACT code, are the subject of further work in preparation.

## 2. The advantages of tungsten and boron as shield materials

The utility of tungsten and boron as shield materials arises from their favourable neutrons cross sections as illustrated in figure 1. Tungsten has several stable isotopes: 30.64% $^{184}$W, 28.43% $^{186}$W, 26.50% $^{182}$W, 14.31% $^{183}$W and 0.12% $^{180}$W. They have similar nuclear properties and figure 1 shows the neutron cross section for natural tungsten. The tungsten cross section is dominated by elastic scattering, but between 10 and 20 MeV there is an appreciable (n,2n) cross section. Between 1 and 8 MeV the inelastic, the (n,n'γ), cross section is appreciable, absorbing much of the neutron energy with the emission of gamma rays. At lower energies, from a few eV to 1 MeV, the tungsten isotopes have many resonances for elastic and (n, γ) reactions. The occurrence of the resonance peaks at differing energies for each isotope contributes to the effectiveness of natural tungsten as a neutron shield.

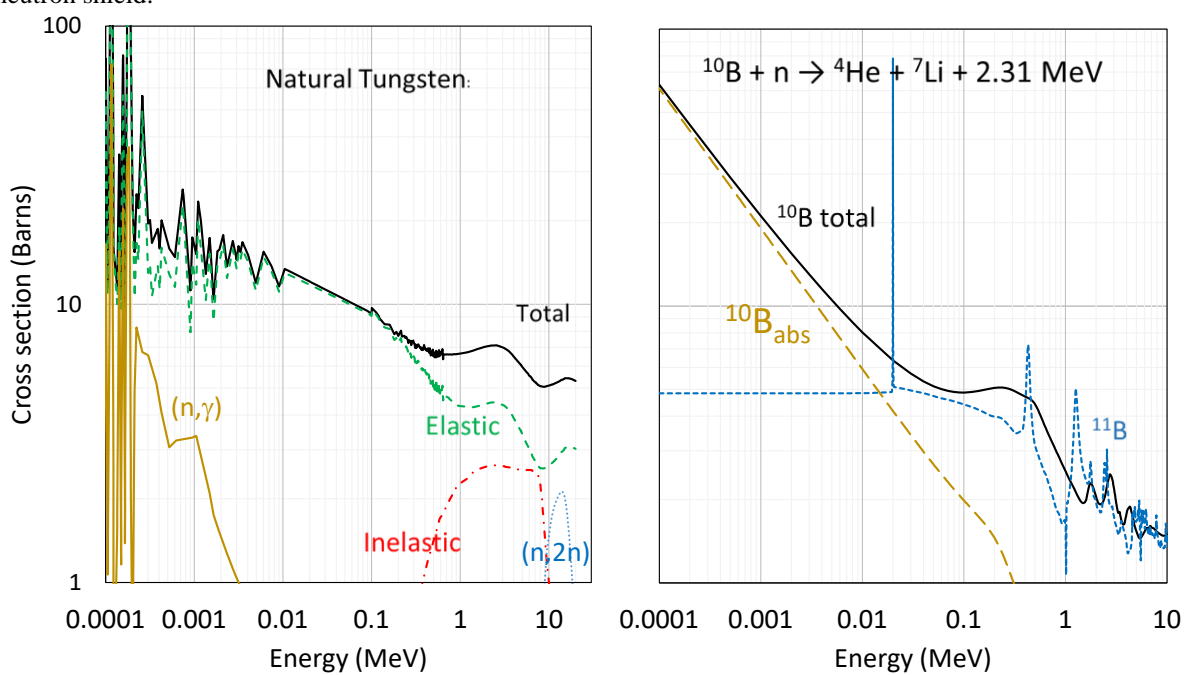

**Figure 1**. The neutron cross sections for tungsten (left) and for boron (right). For tungsten the total (black), elastic (green, dashed), inelastic (red, dash-dot), (n,2n) (blue, dotted) and (n,γ) (yellow) cross sections are shown. For boron the minority (20%) $^{10}$B isotope (black) and the majority $^{11}$B isotope cross section (blue dashed) are shown. Below 10 keV the $^{10}$B cross section follows an inverse velocity increase with decreasing neutron energy as show in the curve labelled $^{10}$B$_{abs}$. The cross sections are from the Brookhaven National Nuclear Data Centre [15].

Boron has two stable isotopes of which the $^{10}$B isotope, with a 20% natural abundance, has the higher neutron absorption coefficient, particularly below 10 keV, as illustrated to the right of figure 1. The $^{10}$B neutron capture gives rise to a $^{4}$He alpha particle, a $^{7}$Li atom and a 0.478 MeV gamma ray. This transmutation means that the $^{10}$B will be gradually be depleted during operations, as will be reported in a later paper. The sub-MeV gamma ray produced has the potentially deleterious consequences of contributing to the power deposition in the HTS core, causing radiation damage to the shield and core, and generally adding to the radiation dose levels produced by the tokamak during operations. There will also be changes in the thermal, mechanical, and corrosion performance in the shield resulting from the transmutation. Isotopically concentrated $^{10}$B is widely used in the nuclear fission industry for neutron-absorbing control rods. For this reason, this study compares the shield performance of natural 20% abundance with shields made from 0, 40, 60, 80 and 100% isotopically enhanced fractions of $^{10}$B.

The properties of tungsten and boron detailed above provide the key to their use in a fusion reactor neutron shield, which must reduce the transmitted power to acceptable levels and limit radiation damage. The plasma-facing side of the shield faces a flux of energetic 14.1 MeV neutrons from the deuterium-tritium reaction

$^2$D + $^3$T = $^4$He + n +17.6 MeV. The 3.5 MeV $^4$He particles are not a problem as their charge keeps most of them within the plasma. The shield can reduce the power transmitted by several methods, which are illustrated in figure S1 in the supplementary material.

(i) Elastic scattering allows many neutrons to be simply reflected back into the fusion chamber. Tungsten atoms make a good reflector because of their high mass compared to a neutron ensures little recoil energy and ~3 barns of elastic cross section per atom at 14 MeV.

(ii) When incident neutrons collide with lighter atoms, like boron, the collisions exchange, or moderate, the neutron energy so reducing the transmitted power. The best moderator is hydrogen since its mass is the same as the neutron and the maximum energy is lost on collision. It is for this reason that water is often used within a shield, where it can also provide cooling, although its oxygen activation may give a problem. Boron or carbon with masses around 10 to 12 provide intermediate moderation.

(iii) Inelastic (n,2n) reactions produce a different isotope of the same element. A typical reaction would be $^{183}$W + n = $^{182}$W + 2n + γ, which is highly beneficial transforming each high energy neutron into two lower-energy neutrons of a few MeV which are easier to shield.

(iv) Inelastic (n,n'γ) neutron scattering means that the incident neutron forms a new compound nucleus which quickly decays to release the neutron with an appreciably lower energy and with the emission of the excess energy in the form of a gamma ray. The neutron energy is reduced by around 1 to 7 MeV, but this energy is released as a gamma ray of comparable energy. The transmitted neutron energy is therefore appreciably reduced but there will be a new gamma flux leading to its own transmitted energy and needing extra gamma shielding.

(v) (n, γ) capture is common at lower neutron energies as indicated in figure 1. The neutron is absorbed to form a tungsten isotope with one higher atomic weight with the emission of a gamma ray of significant energy,

(vi) The lower energy neutrons produced by moderation or inelastic scattering become increasingly likely to be absorbed by isotopes such as $^{10}$B with a high neutron absorption cross section which normally increases with decreasing energy as 1/v where v is the neutron velocity, as illustrated in figure 1. The neutron transmitted power is reduced but the absorption produces a gamma ray leading to increased transmitted gamma power.

Gamma rays are shielded by scattering from electrons in the shield atoms, and thus need high atomic number (number of electrons per atom) and high density (atoms/m$^3$). Tungsten is therefore a comparatively good element for a gamma shield, while boron is not.

Table 1 shows the shielding materials selected for this study. Monolithic tungsten and tungsten carbide are well known shield materials and are included for comparison. Also included is a reactively sintered iron-tungsten borocarbide B$_{0.329}$C$_{0.074}$Cr$_{0.024}$Fe$_{0.274}$W$_{0.299}$ [7], hereafter referred to as the borocarbide. The other materials have been ordered according to their boron atomic fraction. The assumed densities for natural boron-containing materials (20% $^{10}$B isotopic ratio), shown in heavy type, were taken from the literature [10]. The values in the table for other isotopic concentrations have been calculated from the 20% values. To correspond to the densities used in this paper they have been multiplied by 0.98 to allow for extra porosity during bulk manufacture. Since this and other work shows that the shield performance is very sensitive to material densities, the densities of tungsten borides were compared with X-ray powder diffraction measurements found in [16] using samples with the natural 20% $^{10}$B isotopic ratio. It is seen that the assumed densities agree with the X-ray calculated values to within three significant decimals except for WB$_4$ where the assumed value is only 82.5% of the X-ray value. It has been shown that this difference is not sufficient to make any material change to the conclusions of this paper.

In most previous studies the shield design has included annular water-cooling channels [3-5]. Water cooling presents an operational challenge because of the neutron activation reaction $^{16}$O(n,p)$^{16}$N; this nitrogen isotope decays with a half-life of 7.13 s emitting a 6.129 MeV gamma ray. Besides providing cooling, the hydrogen in the water is also an excellent neutron moderator, and this may be a key feature of its success as a shield material. This study attempts to address this point by directly comparing monolithic shields with water-moderated versions containing 5 radial layers of water distributed over 25 radial layers of shield.

**Table 1.** Properties of the boron containing materials considered in this paper along with monolithic tungsten and tungsten carbide included for comparison purposes. Theoretical densities are taken from Ref. [10]. The densities are at 98% of the theoretical values, based on data on W$_2$B-based materials in Ref. [11], which included 2% porosity. The last column gives densities measured by X-rays for this paper [16], multiplied by 0.98.

| Material | Formula | Boron fraction | Average atomic mass (amu) | Density (Mg/m$^3$) | | | | | | X-ray density *0.98 (Mg m$^{-3}$) |
|---|---|---|---|---|---|---|---|---|---|---|
| $^{10}$B fraction | | | | 0% | 20% | 40% | 60% | 80% | 100% | 20% |
| Monolithic tungsten | W | 0.000 | 183.84 | 18.91 | - | - | - | - | - | |
| Tungsten carbide | WC | 0.000 | 97.93 | 15.32 | - | - | - | - | - | - |
| Reactively sintered boride | $B_{0.33}C_{0.07}Cr_{0.02}Fe_{0.27}W_{0.30}$ | 0.329 | 21.97 | 12.38 | **12.37** | 12.36 | 12.35 | 12.35 | 12.35 | - |
| Di-tungsten boride | $W_2B$ | 0.333 | 126.16 | 16.76 | **16.75** | 16.75 | 16.73 | 16.70 | 16.67 | 16.77 |
| Tungsten boride | WB | 0.500 | 97.33 | 15.44 | **15.43** | 15.41 | 15.39 | 15.38 | 15.38 | 15.43 |
| Di-tungsten penta-boride | $W_2B_5$ | 0.714 | 60.25 | 12.94 | **12.91** | 12.88 | 12.85 | 12.81 | 12.78 | 12.72 |
| Tungsten tetra-boride | $WB_4$ | 0.800 | 45.42 | 8.26 | **8.23** | 8.2 | 8.17 | 8.14 | 8.12 | 9.97 |

The attenuating effectiveness of a shield depends on its constituent atomic densities. Here the tungsten density is key to gamma attenuation, the boride (or carbide) to neutron moderation, and the boride to neutron absorption. Figure 2 shows these atomic densities for tungsten and for boron or carbon, along with their sum plotted against boron (or carbon) density. $W_2B_5$ is seen to lie above the trend line in all cases and to give the highest total atomic density.

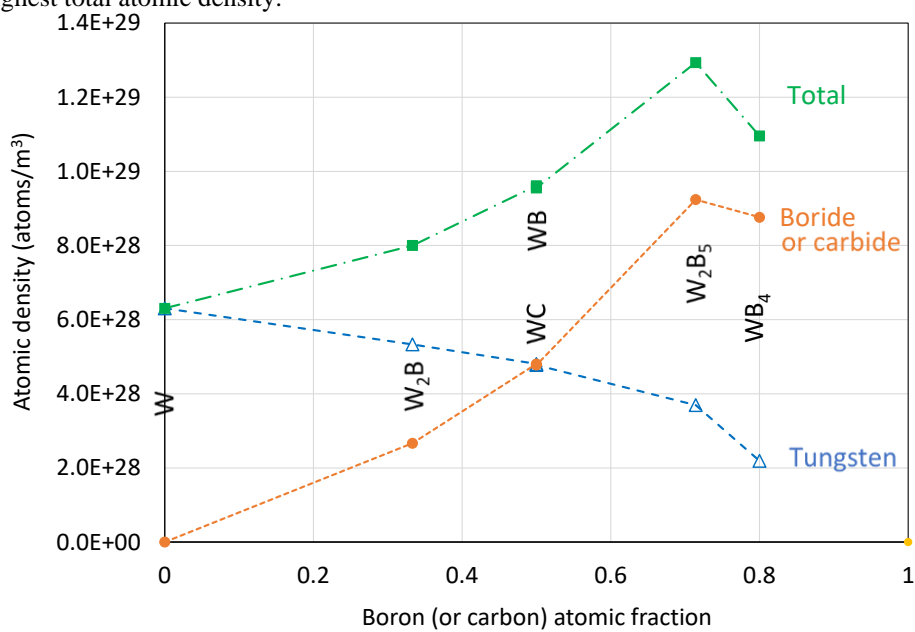

**Figure 2.** The atomic densities within the range of tungsten borides considered for tungsten (blue open triangles), for boride (or carbide) (brown circles) and in total (green squares). For both atomic densities, $W_2B_5$ lies above the trend lines shown. Density data are taken from Ref. [10].

To reduce core power deposition and shield and core irradiation, the main option available is to increase the shield thickness, and consequently the tokamak major radius. In this study some 5 plasma major radii have been chosen, spaced from 1400 to 2200 mm. The superconducting core has been kept constant at 250.9 mm radius. This is consistent with the plasma properties, however in practice it is likely that engineering considerations such as the hoop stress at the top and bottom of the toroidal field coils will determine a modest increase in core size. The shield thickness is adjusted to keep the plasma gap between the inner plasma boundary and the first wall constant. This gave 5 shield thicknesses between 253 and 670 mm.

To summarize the shield materials options of this study, some 7 materials, the 5 boron containing materials each with 6 isotopic concentrations, both with and without water cooling, and with 5 major radii, are considered giving (5x6+2)x5x2=320 shield models. Figure 3 illustrates the radial build of the largest and smallest models.

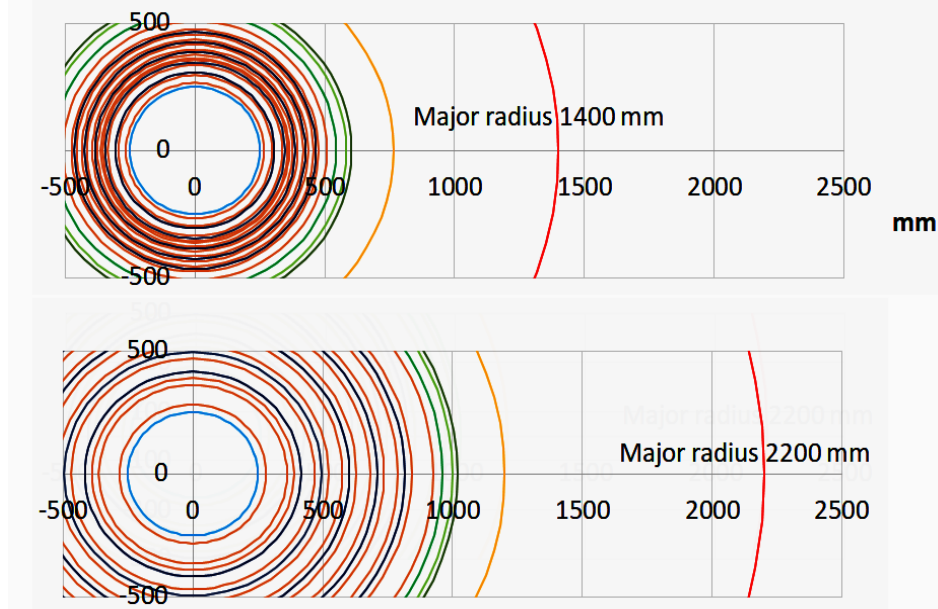

**Figure 3.** The radial build of the smallest and largest tokamak major radii considered. The central core outer radius (blue) is fixed in size as are the thicknesses of the vacuum vessel (light green), the gap between the first wall and the plasma boundary (dark green) and the tungsten first wall (yellow). Dark red shows the shield outer radii, and the dark blue those of the water layers. The plasma central position is shown in bright red.

### 3. Radiation transport modelling

Neutron and photon transport calculations were performed using the code MCNP 6.2.0 [9] invoking the FENDL 3.0 (neutron), MCPLIB84 (photon), and ENDF7U (photo-neutron) cross-section libraries. Calculations were performed in a series of steps:

Firstly, weight windows were generated iteratively for each of the models using the automated weight window generator (WWG) in MCNP using a tally optimized for reducing the variance of low and epithermal energy neutrons and secondary gamma photons in the centre of the core.

Secondly, heat deposition tallies (type F6) were scored within the HTS core for both neutrons and photons using the weight windows generated in the first stage calculation.

Thirdly, neutron only transport models were run (in MODE N) with energy-resolved neutron flux tallies (type F4) using the CCFE-709 energy group scheme and the variance reduction from stage one turned on; the flux tallies were recorded within the HTS core and as a function of shielding depth. In this case, the contribution of photo-neutrons to the neutron flux tallies were assessed for certain cases and deemed insignificant.

The energy depositions and fluxes were computed for a nominal 200 MW fusion power plant. The fusion neutron creation matrix as a function of radius and height was provided for the 1400 mm major radius case from the Tokamak Energy System Code as a matrix of 68 radial segments and 200 vertical segments. Finally, the neutron flux files were input to the activation and transmutation code FISPACT, where dpa, gas rate and $^{10}B(n,\alpha)^{7}Li$ depletion were tracked over a continuous 30-year irradiation period. The results of the FISPACT analysis will be presented in a future publication.

Most of the results presented will correspond to tallies covering the integral over all energies of the full distributions. In particular they are calculated for the following tallies:
1. HTS peak neutron fluence at the vertical mid-plane section in $cm^{-2}MWh^{-1}$ units and its error

2. HTS mean neutron fluence in cm$^{-2}$MWh$^{-1}$ units and its error
3. Total power in HTS in kW and its error
4-29. Neutron fluence for the 25 layers of the shield (cm$^{-2}$MWh$^{-1}$) and error
30. Neutron power in the HTS in kW with error
31. Photon power deposition in the HTS in kW with error.

The table of these 31 parameters for all the 320 calculations is given in the supplementary data. Computations for the single case of natural $W_2B_5$ at major radius 2200 mm were performed with much improved statistics.

### 4. Comparison of monolithic boride material shields

The neutron and gamma power deposition into shields of the monolithic materials with natural boron isotopic concentration, integrated over particle energy, are shown for the monolithic materials at major radii $R_0$=1400 to 2200 mm in figure 4. It is seen that the gamma power deposition exceeds the neutron deposition by around an order of magnitude. This is despite the fact that the radiation incident on the first wall comprises mainly 14.1 MeV fusion neutrons with only around 15% proportion of gammas as shown in figure 4 of reference [5]. Reference [5] showed that the mean neutron energy dropped steadily through a typical moderated neutron shield but that the gamma power remained high, leading to the high gamma power deposition. The graphs in figure 4 show that the power deposition for both neutrons and gammas generally decreases with boron content up to $W_2B_5$, where it forms a local trough for all the major radii considered. The relative change in power deposition decreases with increasing device size. For example, at $R_0$=1400 mm $W_2B_5$ shows lower neutron power than W by a factor of 6 or so, while at $R_0$=2200 mm this factor falls to ~3. It may be concluded that boron, through its combination of neutron moderating and absorption effectiveness, make it the key element in reducing the power deposition in these materials.

The borocarbide $B_{0.329}C_{.074}Cr_{0.024}Fe_{0.274}W_{0.299}$ has good performance at all major radii but is slightly worse than the trend of the binary materials. This is probably because of the fraction of chromium and iron which have less favourable gamma shielding effectiveness than tungsten.

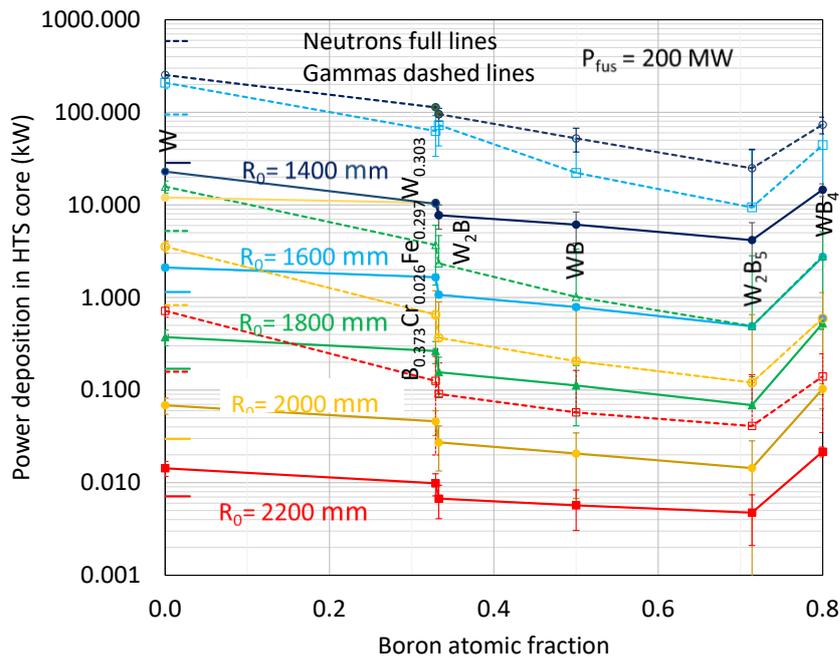

**Figure 4**. The power deposition into the HTS core for the monolithic shields at varying major radii and associated shield thickness. Neutron results are shown as full symbols and continuous lines, gamma results as open symbols and dashed lines. WC has zero boron fraction and is shown as an indent. Statistical errors are shown.

A different presentation of the same data is obtained by plotting the power deposition on a log scale against the major radius for each material as shown in figure 5. The neutron power depositions are shown to follow a roughly exponential decrease with shield thickness, corresponding to a half-power distance of 42.2 mm, as shown by the dot-dashed line, fitted to the monolithic tungsten data. Other materials including $W_2B_5$ show a slight upwards curvature suggesting a decreasing performance over W at larger shield thicknesses. The dashed gamma power deposition follows a similar overall decline with shield thickness with the half-power thickness for gammas essentially the same as that for neutrons, although there is a gradient change around 400 mm thickness.

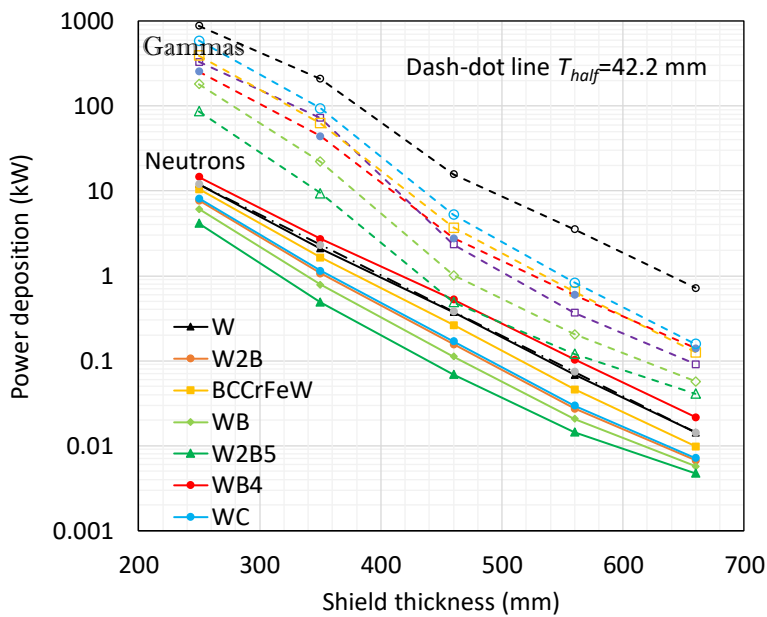

**Figure 5**. The power deposition into the HTS core for neutrons (full lines) and for gammas (dashed) plotted against the shield total thickness. The dash-dot exponential is fitted to tungsten neutron results.

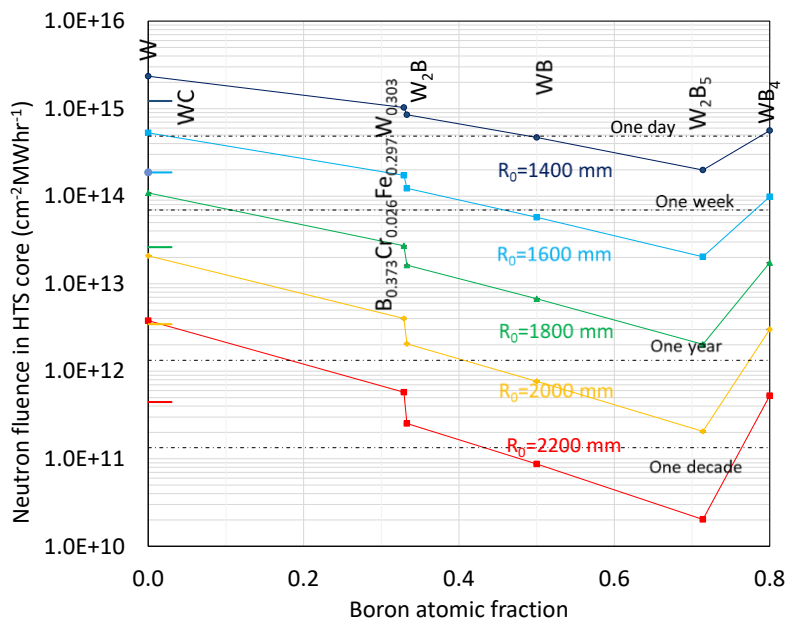

**Figure 6.** The neutron fluence within the HTS core for monolithic material shields at varying major radii. WC has zero boron fraction and is shown by indents. Dash-dot lines show the estimated lifetime of the HTS material from $2 \times 10^{18}$ cm$^{-2}$ neutron fluence [14]. Error bars are included but are less than 1% and hardly visible.

Another important result is the neutron flux within the HTS core. In this paper the neutron core flux quoted will be that at the vertical midplane height in the tokamak where the flux is largest, although the mean flux over the whole of the core was also calculated and is given in the supplementary data. This peak flux is important in determining the lifetime of the HTS core material through its radiation damage. Figure 6 shows the neutron fluence within the HTS core for all major radii again as a function of boron concentration. Also included are estimations for the maximum possible lifetime for a given HTS material, based on HTS performance degradation after neutron fluences of $2 \times 10^{18}$ cm$^{-2}$ [14] at 200 MW fusion power. For example, to enable a decade lifetime, the $R_0$ must be at least 1800 mm, where $W_2B_5$ is the only viable candidate shield material.

The increases in fluence over the 25 tally radii within the shield from the core-side to the plasma-facing side are shown in figure 7 for the major radii of 1400 mm and 2200 mm. The fluence is plotted logarithmically for each monolithic shield material as labelled. It is seen that the 643 mm thick shield at 2200 mm major radius covers some 5 orders of magnitude compared to less than two for the 253 mm thick shield at 1400 mm radius. In both cases $W_2B_5$ performs significantly better than the other materials. The order of performance with material is slightly changed for the smaller major radius, with the WB and $W_2B$ performing relatively better than $WB_4$ at the larger device size. Near the plasma-facing edge all the materials face a similar incident fluence and the differences between materials are relatively low.

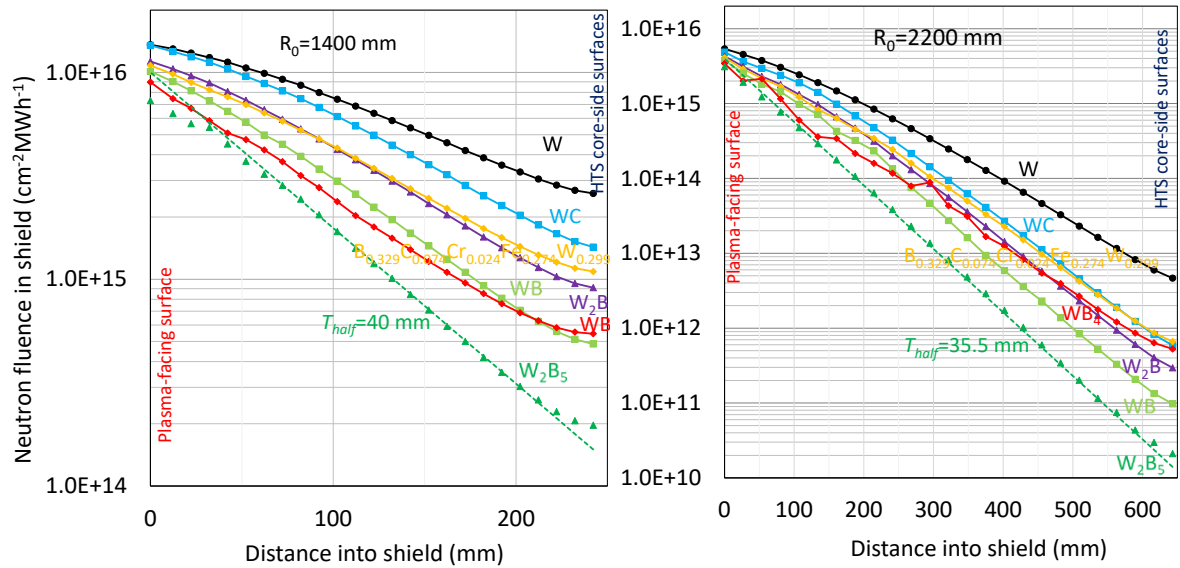

**Figure 7.** The neutron fluence for monolithic material shields at $R_0$ =1400 mm (left) and 2200 mm (right) plotted against distance into the shield. Note the scale changes. The points for $W_2B_5$ are shown against the dashed exponential decrease with the half-attenuation distances shown. The plot for $W_2B_5$ at $R_0$ = 2200 mm has improved statistics.

All the plots in figure 7 show a similar shape with a gradient build up from the plasma-facing side to a close to exponential decrease in the centre of the shield, A similar but lesser decrease in gradient occurs close to the HTS core side of the shield. This behaviour is illustrated by the dashed lines in figure 7 which show an exponential decrease in fluence with the half attenuation distances shown. The deviations from the dashed lines are clearly seen for $R_0$ =1400 mm, although for $R_0$ =2200 mm they are quite small. It is probable that these deviations are partly caused by the lack of absorbing boron in the central superconducting core, and at the stainless steel and tungsten plasma-facing components as illustrated in figure 3. Although the scales of the plots at the two radii are very different, the half attenuation distances of the dashed lines varies only slightly.

For the case of a shield of monolithic $W_2B_5$ at 2200 mm major radius results of higher accuracy were available. The top left-hand of figure 8 shows the neutron energy lethargy spectrum as a function of distance into the shield for this case. Lethargy spectra are appropriate when covering the large energy range from 1 keV to over 10 MeV. In a lethargy spectrum the quantity plotted per energy bin is multiplied by the bin mean energy and divided by the bin width in energy, so that the quantity plotted is dimensionless in energy and sums to the total quantity. The spectra has been smoothed by averaging of 3 to 5 energy bins for the energies below 0.02 MeV and for distances into the shield above 500 mm.

A striking characteristic of the top left-hand side of figure 8 is that, with the exception of the red plasma-facing spectrum, the shapes of the energy spectra up to around 6 MeV are remarkably constant. This feature is revealed in the top right-hand side of figure 8, where the intensity has been scaled by the integral of the lethargy fluence up to 6.02 MeV. There is a remarkable superposition of the fluence energy spectra from the majority of the layers within the shield at energies below 6.02 MeV. The surprising conclusion is that the neutron energy spectra below 6 MeV are independent of the depth into the shield!

While this superposition of the spectra occurs up to 6 MeV, at higher energies the situation is quite different. The detail of the scaled lethargy fluence at energies above 6 MeV is revealed in the lower half of the figure which shows the same scaled fluence between energies of 1.5 and 15 MeV (bottom left) and between 12.3 and 14.1 MeV (bottom right). The data in this region can be approximated by the exponential decay given by the dashed line described by $\log_{10}[F(E)/F(6.02)] = -1.9\log_{10}[E/6.02]$ where $F(E)$ is the fluence at energy $E$ MeV.

The strong superposition of scaled fluences below 6.02 MeV, and lack of superposition above this energy suggests quite distinct origins for the flux distributions above and below the black dashed 'dividing line' shown in the lower left figure. It is suggested that the fluence below this dividing line is caused by 'inelastic' scattering processes such as the (n,2n) and (n,n'γ) cross sections described in section 2 which reduce neutrons of energy around 14 MeV to a broad inelastic spectrum of neutron energies centred around 0.2 MeV and remarkably constant at all distances into the shield. In contrast the bottom right plot in figure 8 shows that there is little superposition above 6.02 MeV and above the dividing line. This 'elastic' fluence is composed of the original 14.2 MeV fusion neutrons which may have lost energy principally by elastic scattering from the tungsten and boron atoms as they pass through the shield as detailed in section 2. With each increment into the shield, the 14.1 MeV scaled fluence decreases. At distances into the shield above 300 mm, a peak in the scaled fluence appears below 14.1 MeV and shifts to ever lower energies.

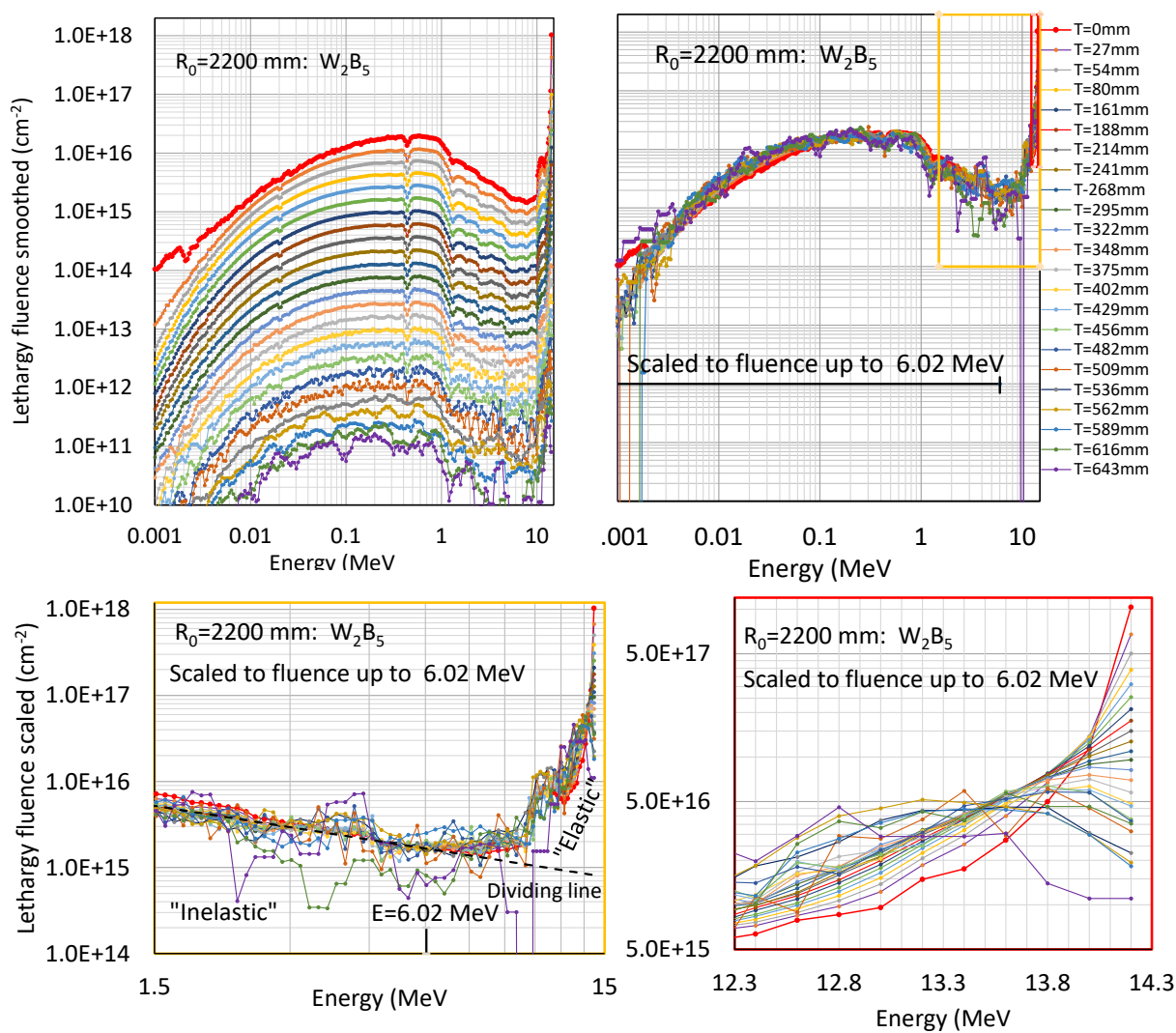

**Figure 8.** Top left: the smoothed lethargy spectrum of the neutron fluence within the shield for $W_2B_5$ at $R_0$=2200 mm plotted as a function of energy for the various distances into the shield. Top right is the data scaled according to the total fluence at energies up to 6.02 MeV. Bottom left (yellow outline) is the scaled data in the energy range above 1.5 MeV. The dashed line is used to divide the total fluence spectrum into "elastic" and "inelastic" energy spectra above 6.02 MeV. Bottom right (red outline) shows the broadening of the elastic scattering component near 14 MeV.

Making this assumption, it is straightforward to sum the elastic and inelastic components of the original unscaled fluence at each distance into the shield as shown in figure 9. It is seen that both components start with a similar fluence, and both decay approximately exponentially with distance into the shield as shown by the dashed lines, with the corresponding half-distance lengths shown. The elastic component is closest to exponential with a half-attenuation distance of 33 mm. The inelastic component has lesser decay rates near both plasma-facing and core-facing sides decay as was noted for the total fluences through the shield in figure 7 again reflecting the relative lack of absorbing boron material at these points. The result is that the ratio of the elastic to the inelastic fluence decreases with distance into the shield from about 1 at the plasma-facing side to 5 at the core-facing side. The elastic component has a slowly decreasing mean energy from 14.0 MeV at the plasma-facing side to 12.8 MeV at the HTS-facing side. The inelastic component mean energy slowly decreases from 0.8 MeV at the plasma-facing side to 0.7 MeV at the HTS-facing side.

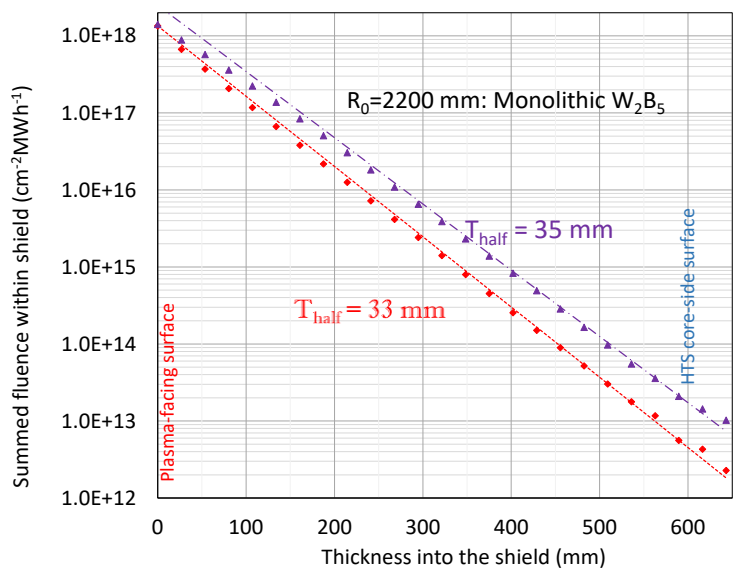

**Figure 9.** The "elastic" and "inelastic" components of the neutron fluence summed over all energies as a function of distance into the shield for $W_2B_5$ at $R_0=2200$ mm. The dashed lines are exponential fits over the central region of the shield.

The scaled superposition method worked well for $W_2B_5$ at other radii, and for $WB_4$, but not for other shield materials. Figure 10 shows the unscaled neutron fluence energy spectra for tungsten and tungsten boride materials with increasing boron content. The loss of superposition is seen by changes in the energy spectrum profile and decreases inversely as the boron concentration. Superposition works well in the range 1 to 6 MeV just to the left of the minima in the spectra, but then fails increasing badly as the energy decreases. There are much greater fluences at the core-side of the shield than would have been expected from superposition, as indicated by the violet circles in the figure.

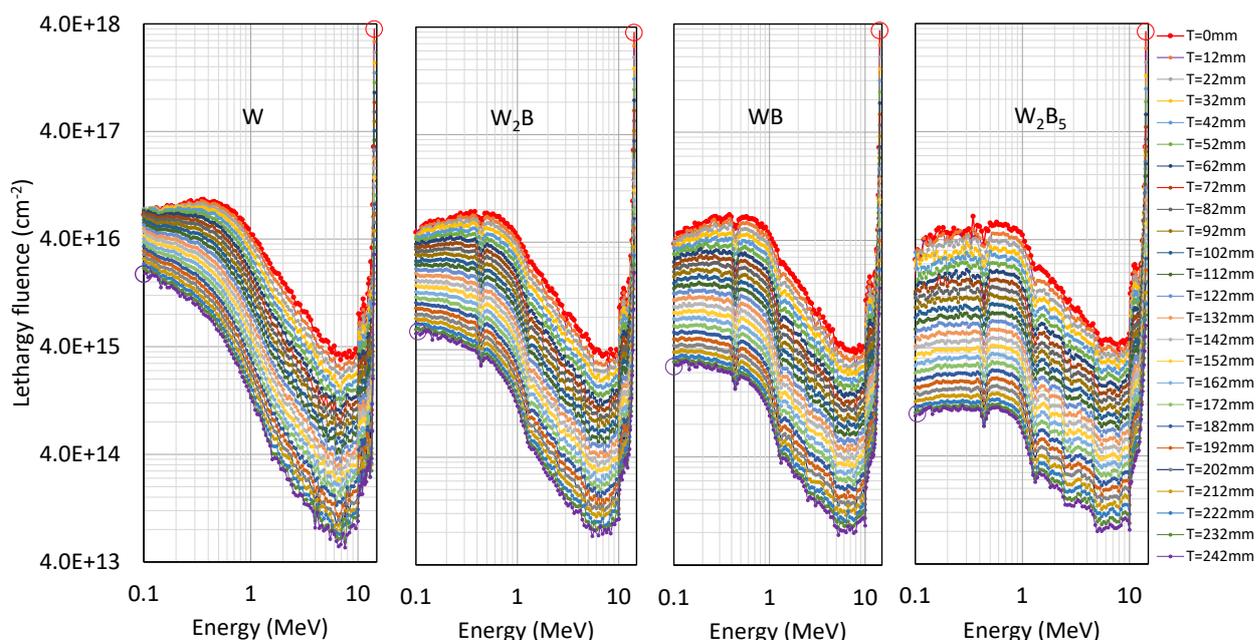

**Figure 10.** The energy spectrum through the shield from 0.1 to 14.1 MeV for several of the shield materials. The 25 layers are from the plasma-facing layer (red) to the HTS-facing layer (violet). Note the almost constant peak at 14.1 MeV (red circles), the very similar fluences around the fluence minima at 6 MeV and the decrease of the HTS-facing fluence at 0.1 MeV, with increasing boron content (violet circles).

The constant neutron energy spectrum through the shield observed for the boron concentration of $W_2B_5$ suggests a rationale for its exceptional performance. As the depth into the shield increases, an ever-decreasing quantity of the broad 'inelastic' neutron spectrum will be added in proportion to the remaining 'elastic' fluence of around 14 MeV fusion neutrons. At the same time the spectrum will be softened to lower energies by moderation. For the total spectrum profile to be maintained, the $^{10}B$ isotope absorption must be sufficient to remove the lower energy neutrons so produced. In contrast, with for example a $W_2B$ shield, the $^{10}B$ absorption is insufficient for this, and the neutron spectrum at lower energies builds up continuously with distance into the shield as suggested by figure 10 leading to an overall larger power deposition.

The energy spectra per unit lethargy of the deposited power into the superconducting core for the monolithic materials as shown in figure 11 for major radii of 1400 mm (left) and 2200 mm (right). It is striking that the gamma fluence is ~100 higher than the neutron fluence above 0.1 MeV but drops rapidly below 1 keV as its attenuation rapidly increases. Most of the power deposition into the core is within the 0.1 to 10 MeV range for both gammas and neutrons. The order of materials is maintained over most of the energy range, for both radii, and for both neutrons and gammas. Comparison between the attenuations at the two major radii show a factor of ~100 for gammas above 1 MeV, but a larger ~1000 factor for neutrons which is roughly maintained over the whole energy range. An exception is the energy range 1-14 MeV where $WB_4$ has the highest power deposition. The atomic fraction of boron could be critical in this energy range where figure 1 shows the boron cross section to be low.

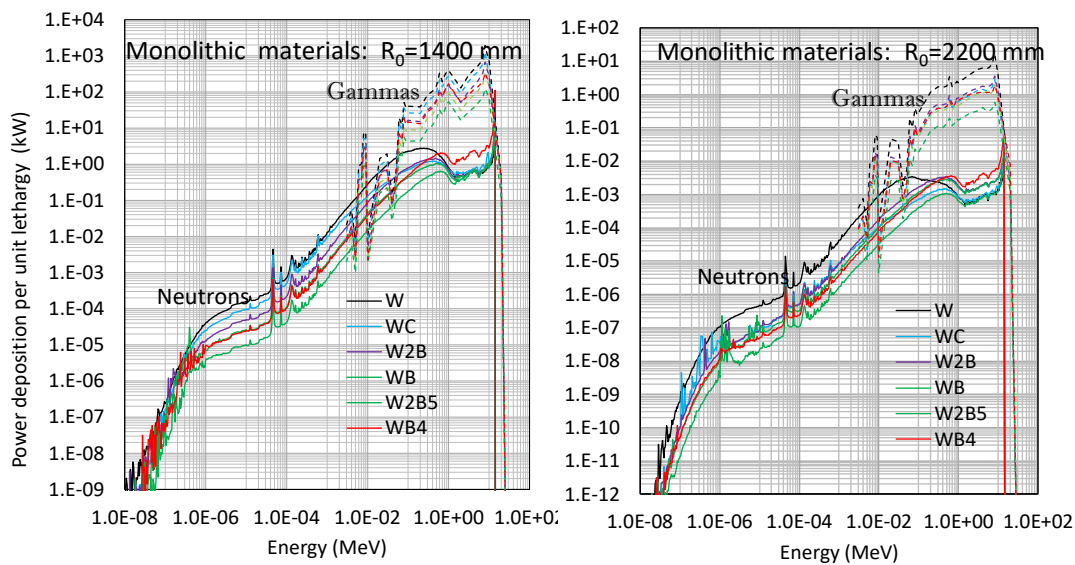

**Figure 11.** Power energy spectra per unit lethargy of the neutron and gamma power deposition into the superconducting core for monolithic material shields for the cases of $R_0$ =1400 mm (left) and $R_0$ =2200 mm (right). Full lines are neutrons, dashed are gammas. Note the very different scales for the two major radii.

### 5. Results with water-cooled shields

For each of the chosen shield materials, and for each boron concentration, the effects of introducing a water moderator were evaluated. The thickness of the water layers varied from 10 mm at $R_0$ =1400 mm to 17 mm at $R_0$ =2200 mm. Pure water has been previously chosen as moderator [3-5] but because of its activation problem it may be desirable to avoid it. A fixed geometry for the water layers has been used with 5 radii spaced through the shield as was indicated in figure 3.

Figure 12 shows the neutron and gamma power depositions into the HTS core, with and without water layers within the shield materials for major radii of 1400 mm (left) and 2200 mm (right). The presence of water layers tends to flatten out the variation of deposited power with boron fraction with decreases at high depositions and increases at low depositions. For tungsten and WC, the reduction is considerable. $W_2B_5$ maintains its optimal position with the lowest deposited power at all major radii, and with this material, there is always a detrimental effect from the addition of water layers. In the larger radius device, the detrimental effect of water extends to all of the pure tungsten boride materials.

Similar results were obtained for the neutron fluence in the HTS core and are shown in the Supplementary Material figure S3. They show that including water layers smooths out the boron fraction dependence with a lower fluence at the pure tungsten end and higher fluence for $W_2B_5$.

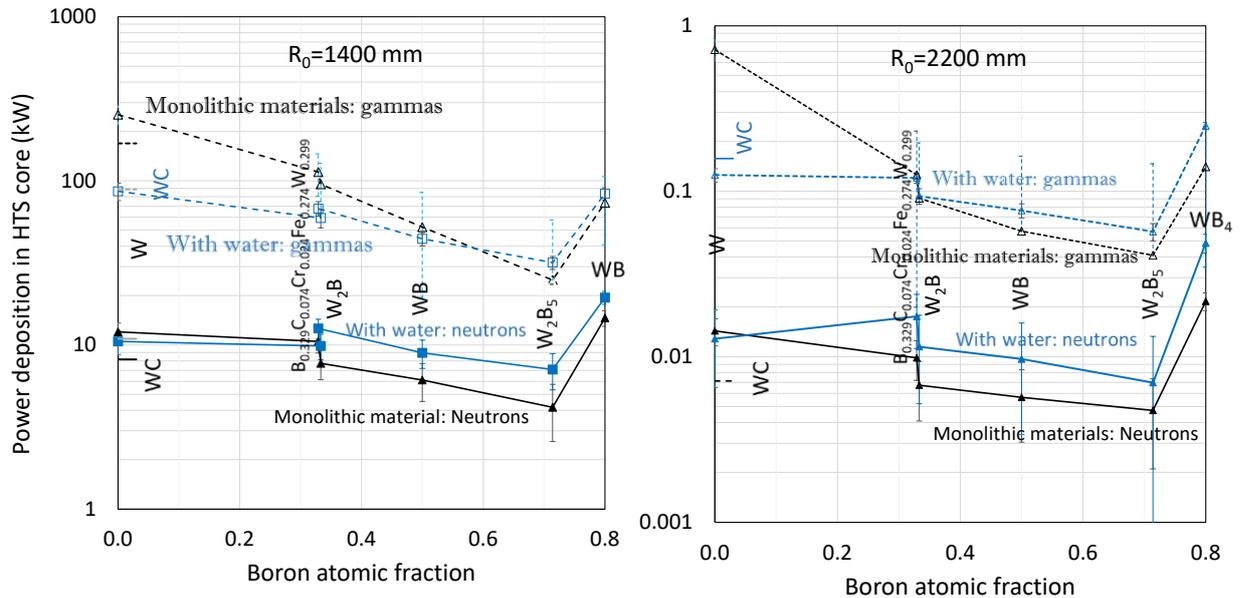

**Figure 12.** The neutron (full) and gamma (dashed) power deposition into the HTS core for monolithic materials (black) compared with those with water shielding (blue) for major radii of 1400 mm (left) and for 2200 mm (right).

The energy spectrum of the power deposition is shown in figure 13 for monolithic $W_2B_5$ shields and those with water moderation at major radius 1400 mm (left) and at 2200 mm (right). It is seen that the effects of moderation are almost negligible except at the two energy extremes. At energies from 1 to 14 MeV, the presence of a moderator increases the power deposition by around a factor of 2. The 14 MeV peak in the neutron spectrum even at the HTS core is clear from the figure. These are scattered by water producing increased numbers of MeV neutrons. At the lowest energies below 10 eV the boron in the shield preferentially removes these neutrons and causes the reduction in neutron power deposition

The gamma spectrum relates closely to the neutron one. $(n,\gamma)$ and $(n,n'\gamma)$ reactions in the tungsten atoms increase the gamma spectrum, in proportion to the neutron spectrum. Below 1 MeV, when the neutron spectrum is not dependent on the presence of moderator, the gamma spectrum is similarly independent.

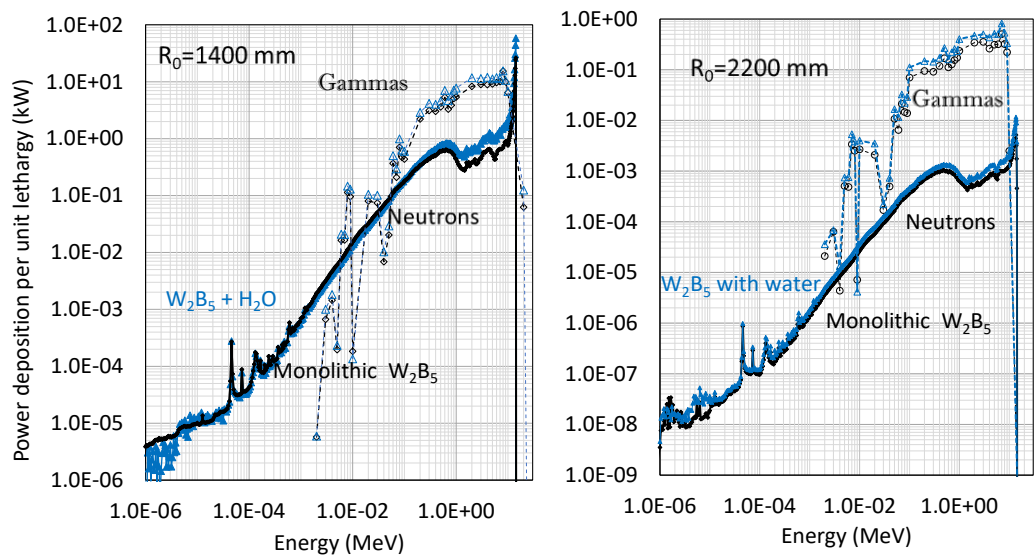

**Figure 13.** The energy spectra at constant lethargy of the power deposition arising from neutrons (full) and gammas (dashed) in the HTS mid-plane region for $W_2B_5$ shields of monolithic materials (black) and for those including water layers (blue) at $R_0$=1400 mm (left) and at $R_0$ =2200 mm (right). Note the change of vertical scale.

Another factor suggested previously [4] was that the tungsten carbide and water shield could be much improved by just a single layer of borated material placed on the inside HTS facing layer of the shield where the neutron energies are lower, and neutron absorption stronger. To check this, calculations were performed to find the HTS power deposition with iron-tungsten borocarbide $B_{0.329}C_{0.074}Cr_{0.024}Fe_{0.274}W_{0.299}$ placed in this position within a WC and water layered shield. The results are shown in table 2 which confirms the appreciable reduction of the borated inner layer shield compared with the WC and water shield. The $W_2B_5$ shields with and without water continue to have superior power depositions.

**Table 2.** The power depositions in kW for selected shields at the chosen major radii.

| Shield Composition | Shield thickness (mm) | | | | |
|---|---|---|---|---|---|
| | 253 | 357 | 462 | 566 | 671 |
| WC+H$_2$O | 99.7 ± 5% | 17.4 ± 3% | 3.07 ± 3% | 0.628 ± 3% | 0.144 ± 4% |
| WC+H$_2$O. Final WC layer $B_{0.33}C_{0.07}Cr_{0.02}Fe_{0.27}W_{0.30}$ | 76.4 ± 2% | 12.9 ± 1% | 2.40 ± 1% | 0.509 ± 2% | 0.118 ± 3% |
| W$_2$B$_5$+H$_2$O | 38.8 ± 3% | 6.14 ± 2% | 1.08 ± 4% | 0.210 ± 3% | 0.064 ± 5% |
| Monolithic W$_2$B$_5$ | 29.3 ± 3% | 3.58 ± 3% | 0.561 ± 4% | 0.135 ± 4% | 0.046 ± 6% |

## 6. The effects of $^{10}$B isotope concentration

The results so far have been for the 20% atomic concentration of the absorbing isotope $^{10}$B in natural boron. It is possible to create shield materials with an enhanced concentration of $^{10}$B for additional cost. The results are shown in figure 14 and in all cases the power reduces with increasing $^{10}$B fraction. For shield materials with a low boron fraction like $W_2B$ it continues to reduce over the whole range. For materials like $WB_4$ with more boron there is more saturation of the effect. Most of the possible gain is achieved by 40% isotopic fraction. It is seen that the power reduction with $^{10}$B isotope is most noticeable for gammas rather than neutrons.

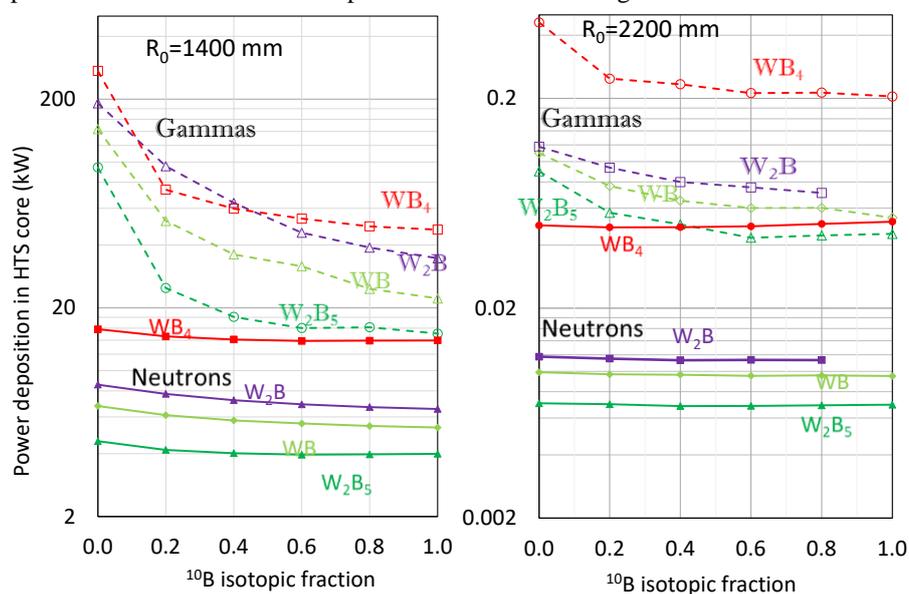

**Figure 14**. The power deposition arising from neutrons (full) and gammas (dashed) in the HTS mid-plane region for monolithic shields at $R_0$ =1400 mm as a function of the fraction of the $^{10}$B isotope. Note the factor 1000 change of scale.

The corresponding results for the neutron fluence as a function of $^{10}$B isotopic fraction shown in figure S5 in the supplementary material are distinctly different. The neutron fluence at both radii falls off much more steeply with isotopic fraction, and there is little of the levelling off seen in the neutron power deposition but rather follow the trend in the gamma power deposition that was shown in figure 14. This suggests that $^{10}$B efficiently captures the lower energy neutrons, but that this effect is slightly off-set by its reduced moderating capability for higher energy neutrons, leading to a lower fluence, but a similar overall neutron power.

Figures for the neutron fluence across $W_2B_5$ shields without and with water content are shown in figure 15 for major radius 1400 mm. The water moderation appears to increase the neutron fluence at all radii except in the case of zero $^{10}$B isotopic concentration. The positions of the water layers are shown by vertical blue lines along the radius axis, and some effects are reflected in the corresponding fluence. The effects of increasing the $^{10}$B isotopic concentration seems to be greater for the monolithic shield than for the water moderated shield.

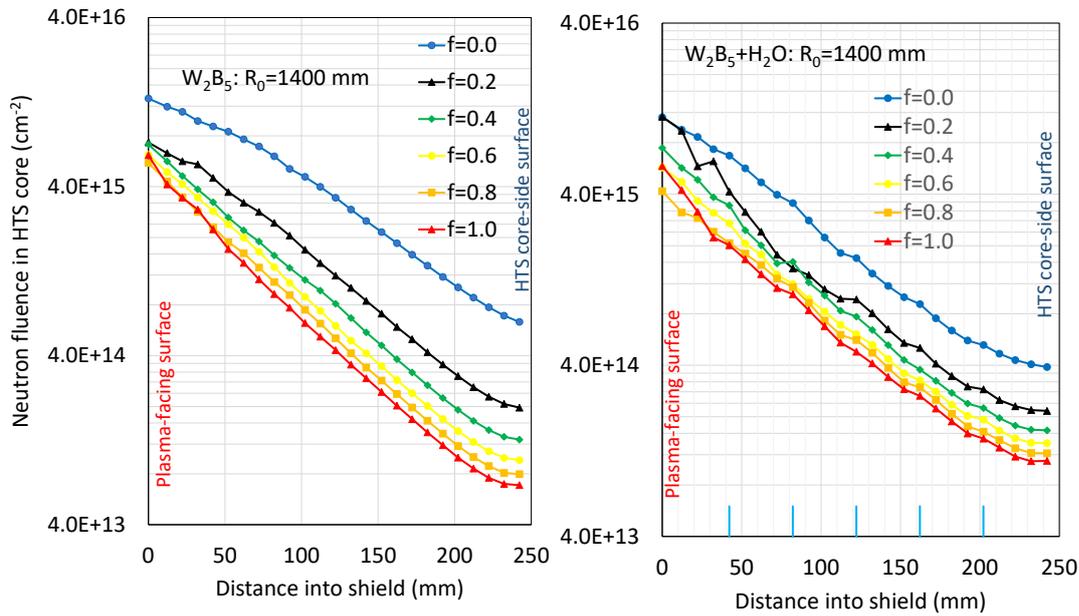

**Figure 15**. The neutron fluence across the shield at major radius $R_0 =1400$ mm as a function of $^{10}$B isotopic concentration for monolithic $W_2B_5$ (left), and for water moderated (right) shields. The blue verticals show water layer positions.

## 7. Thermal strain

Having identified the optimal shielding compounds from an power deposition perspective, we now examine some practical implications of the candidate materials. Table 3 shows their thermophysical properties. The molar heat capacity at room-temperature, $C_m$ is taken from Refs [20-22]. $C_m$ of WB$_x$ compounds decreases with increasing boron content. The values of $C_m$ are converted into the volumetric heat capacity $C_{vol}$ using:

$$C_{vol} = \frac{C_m \cdot \rho \cdot n}{M},$$

where $\rho$ is the density, $M$ is the molar mass and $n$ is the number of atoms in the formula unit. The $\rho$ of the W-B compounds showed notable variation in the literature. For example, as shown in table 1 the $\rho$ of WB$_4$ is given as 8.4 g cm$^{-3}$ by [10] and 10.2 g cm$^{-3}$ by [16], which could be due to variations in sample porosity. To avoid such variability, the values for the borides are selected from only one source reference [10], for consistency. $C_{vol}$ generally shows the opposing trend to $C_m$ with respect to boron content (with the exception of WB$_4$) i.e. $W_2B_5$ has the highest $C_{vol}$ of any of the candidate materials. This result is likely due to the higher atomic packing density in $W_2B_5$ compared to monolithic W as shown in Figure 3.

$C_{vol}$, the heat capacity, can be related to the anticipated thermal strains by defining a simple figure of merit, the expansion per unit energy, $\alpha_v/C_{vol}$ which describes the amount of volumetric expansion per unit of thermal energy absorbed. $\alpha_v$ is taken from refs [18,10,21]. NB: this figure of merit neglects heat exchange with the environment, which was necessary as thermal conductivity data are not available for the tungsten borides.

**Table 3:** Molar Heat Capacity $C_m$, Molar Mass $M$, Theoretical density $\rho$, Volumetric Heat Capacity $C_{vol}$, Volumetric Thermal Expansivity $\alpha_v$ and Expansion per Unit Energy $\alpha_v/C_{vol}$

| Material | $C_m$ (J K$^{-1}$ mol$^{-1}$ at$^{-1}$) | M (g mol$^{-1}$) | $\rho$ (g cm$^{-3}$) | $C_{vol}$ (J cm$^{-3}$ K$^{-1}$) | $\alpha_v$ (10$^{-6}$ K$^{-1}$) | $\alpha_v/C_{vol}$ (10$^{-6}$ cm$^3$ J$^{-1}$) |
|---|---|---|---|---|---|---|
| W | 24.35 [18] | 183.84 | 19.25 [18] | 2.550 | 13.5 [18] | 5.28 |
| $W_2B$ | 23.03 [19] | 378.49 | 17.09 [10] | 3.120 | 20.1 [10] | 6.44 |
| WB | 20.38 [19] | 194.65 | 15.74 [10] | 3.295 | 20.7 [10] | 6.28 |

| | | | | | |
|---|---|---|---|---|---|
| $W_2B_5$ | 15.86 [19] | 421.74 | 13.17 [10] | 3.468 | 23.4 [10] | 6.75 |
| $WB_4$ | 14.01 [19] | 227.08 | 8.40 [10] | 2.591 | 17.4 [10] | 6.71 |
| WC | 17.57 [20] | 195.85 | 15.67 [21] | 2.833 | 16.5 [21] | 5.82 |

Despite $W_2B_5$ having the largest $C_{vol}$, which tends to decrease $\alpha_v/C_{vol}$, this is more than offset by its large $\alpha_v$. Thus, the value of $\alpha_v/C_{vol}$ is greatest for $W_2B_5$. This means it will undergo the highest amount of thermal expansion for every joule of thermal energy absorbed, while pure W will undergo the least expansion. This suggests that thermal strain in $W_2B_5$ may be significantly higher than in W, which may lead to higher thermal stresses.

The high thermal strain of $W_2B_5$ is exacerbated by its very impressive power absorption capability, as shown in the MCNP calculations. Thus, the expansion for a given neutron fluence is expected to be even more stark than the comparisons shown in table 3, suggesting that that the thermal stresses in a $W_2B_5$ shield will be significantly higher than with monolithic W. This underlines the importance of assessing the thermal stress resistance of candidate tungsten boride shields in future work.

## 8. Conclusions

Figure 16 graphically summarises the results of $^{10}B$ enrichment and boron content with a 3-dimensional surface of gamma power deposition as a function of both parameters. The plot illustrates the local minimum power deposition for $W_2B_5$, shown previously in Figures 4, 6 and 12, is maintained at all $^{10}B$ enrichment levels. It also shows that the optimal material, i.e. $W_2B_5$ with 60% $^{10}B$ or more, shows a factor of ~20 lower gamma power deposition compared to monolithic W.

It was previously thought that a hydrogenous moderating material, such as water or a metallic hydride, was an essential part of any tokamak shield, reducing the power deposition and neutron and gamma fluences. This paper shows that an alternative shield employing light atoms such as boron (Z=5) or carbon (Z=6) can be equally as effective, and in the case $W_2B_5$, even more so. Boron introduces a third factor, neutron absorption, as its absorption cross section rises with decreasing neutron energy.

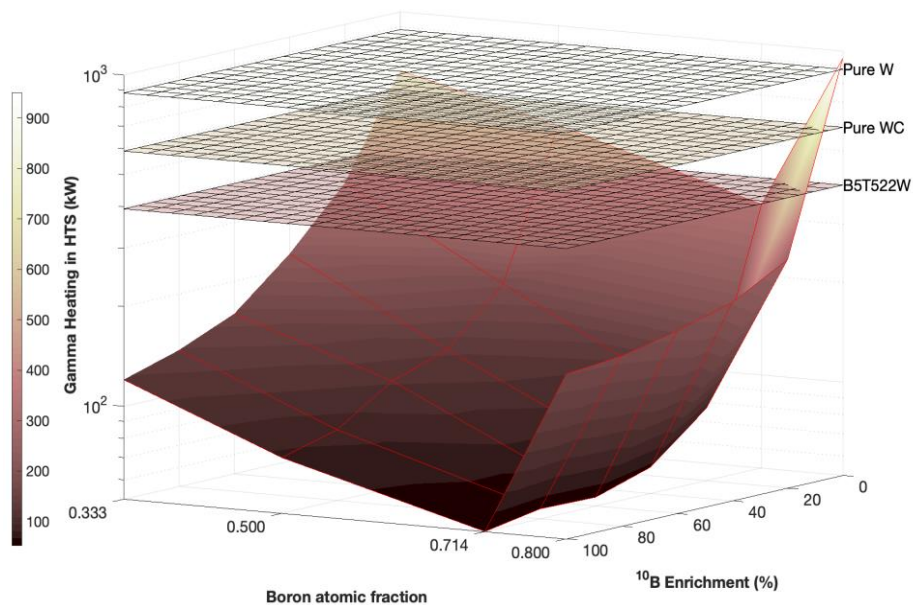

**Figure 16.** The gamma power deposition into the HTS core for the monolithic material shields at $R_0$=1400 mm, as functions of boron atomic fraction and $^{10}B$ isotopic concentration. The cemented boride $B_{0.329}C_{0.074}Cr_{0.024}Fe_{0.274}W_{0.299}$ is shown with natural boron 20% isotopic concentration.

This paper shows that increasing the shield's boron content generally decreases the power deposition to a factor of 10-20 below that of monolithic tungsten, depending on the $^{10}B$ content, suggesting that increasing neutron moderation and absorption within the shield is a key factor. The trend then reverses between $W_2B_5$ and $WB_4$ deposition, probably because of the low density of $WB_4$ and its relative lack of gamma attenuating tungsten. Our study also shows that a monolithic $W_2B_5$ shield far outperforms previously-considered shield architectures composed of outer WC layers and inner borocarbide layers [4].

Energy spectra of the neutron fluence through a $W_2B_5$ shield at $R_0$=2200 mm suggest that, with this boron fraction, the energy spectrum of inelastically scattered neutrons is almost independent of distance into the shield and attenuates through the shield at much the same rate as the elastically scattered fusion neutrons.

The conclusion is that $W_2B_5$ provides an optimal tokamak shield with its good compromise between gamma shielding, neutron moderation and neutron absorption. At the same time, this study also highlights that the candidates with the best shielding capability are also susceptible to the highest thermal strains. Further understanding of their thermal and mechanical properties, and their evolution under radiation damage, is therefore needed. The activation results for the same shield materials and configuration made using the FISPACT code using neutron fluxes from this study will be presented in subsequent work.